\documentclass[runningheads]{llncs}

\usepackage{amsfonts,proof,qtree,amsmath,amssymb,mathtools,algpseudocode,algorithm}
\usepackage{latexsym}
\usepackage{graphicx}
\usepackage[usenames,dvipsnames]{color}
\usepackage{listings}
\usepackage{float}
\usepackage{multirow}
\usepackage{booktabs}
\usepackage{mathrsfs}
\usepackage{mathpartir}
\usepackage{dsfont}
\usepackage{stmaryrd}
\usepackage{url}
\usepackage{textcomp}
\usepackage{hyperref}
\usepackage{alltt}
\usepackage{bbm}
\usepackage{alltt}
\usepackage{xspace}
\usepackage{enumitem}
\usepackage{lipsum}
\usepackage{wrapfig}
\usepackage[usenames,dvipsnames]{xcolor}

\usepackage{cite}
\usepackage{comment}
\usepackage{ebproof}
\usepackage{subfig}
\usepackage{graphics}
\usepackage{tikz}
\usetikzlibrary{arrows,decorations.markings}
\usepackage{enumitem}
\usepackage{hhline}
\usepackage{colortbl}
\usepackage{ebproof}
\usepackage{centernot}
\usetikzlibrary{calc,positioning,math}

\setlist[itemize]{topsep=0pt,itemsep=-1ex,partopsep=1ex,parsep=1ex,itemindent=0pt,leftmargin=8pt}

\definecolor[named]{ACMBlue}{cmyk}{1,0.1,0,0.1}
\definecolor[named]{ACMYellow}{cmyk}{0,0.16,1,0}
\definecolor[named]{ACMOrange}{cmyk}{0,0.42,1,0.01}
\definecolor[named]{ACMRed}{cmyk}{0,0.90,0.86,0}
\definecolor[named]{ACMLightBlue}{cmyk}{0.49,0.01,0,0}
\definecolor[named]{ACMGreen}{cmyk}{0.20,0,1,0.19}
\definecolor[named]{ACMPurple}{cmyk}{0.55,1,0,0.15}
\definecolor[named]{ACMDarkBlue}{cmyk}{1,0.58,0,0.21}
\definecolor[named]{PaleGreen}{RGB}{196, 255, 231}
\definecolor[named]{PaleOrange}{RGB}{255, 213, 169}
\definecolor{intnull}{RGB}{213,229,255}

\makeatletter
\def\arcr{\@arraycr}
\makeatother

\makeatletter
\newcommand{\mtmathitem}{%
\xpatchcmd{\item}{\@inmatherr\item}{\relax\ifmmode$\fi}{}{\errmessage{Patching of \noexpand\item failed}}
\xapptocmd{\@item}{$}{}{\errmessage{appending to \noexpand\@item failed}}}
\makeatother

\definecolor{shadecolor}{gray}{1.00}
\definecolor{ddarkgray}{gray}{0.75}
\definecolor{darkgray}{gray}{0.30}
\definecolor{light-gray}{gray}{0.87}

\newcommand{\hshs}{\mathit{hs}}

\newcommand{\ind}{\mathit{\iota}}
\newcommand{\inds}{\mathit{{\iota}\!{s}}}

\newcommand{\ie}{\emph{i.e.}\xspace}

\newcommand{\eg}{\emph{e.g.}\xspace}

\newcommand{\etal}{\emph{et~al.}\xspace}

\newcommand{\cf}{\textit{cf.}\xspace}
\newcommand{\wrt}{\emph{wrt.}\xspace}
\newcommand{\Iff}{\emph{iff}\xspace}

\newtheorem{pattern}{Pattern}
\newtheorem{prop}{Property}

\newcommand{\mute}[1]{{#1}}

\newcommand{\is}[1]{\mute{\textcolor{ACMBlue}{(Ilya: {#1})}}}
\newcommand{\code}[1]{\lstinline{#1}}
\newcommand{\ccode}[1]{\text{\lstinline[basicstyle=\small\ttfamily]{#1}}}
\newcommand{\cccode}[1]{\text{\emph{\lstinline[basicstyle=\small\ttfamily]{#1}}}}

\newcommand{\set}[1]{\left\{{#1}\right\}}

\newcommand{\asgn}{\leftarrow}

\newcommand{\eqdef}{\triangleq}

\newcommand{\ncr}[2]{\left(\begin{matrix} #1 \\ #2 \end{matrix} \right)}
\newcommand{\stirlingsnd}[2]{\left\{\begin{matrix} #1 \\ #2 \end{matrix} \right\}}
\newcommand{\ordnat}[1]{\mathds{Z}_{#1}}
\newcommand{\bloomfilter}{\mathit{bf}}
\newcommand{\quotientfilter}{\mathit{qf}}

\newcommand{\vseq}{\mathit{vs}}
\newcommand{\yseq}{\mathit{ys}}
\newcommand{\xseq}{\mathit{xs}}
\newcommand{\countingbloomfilter}{c\!f}
\newcommand{\bfquery}[2]{#1 \in_? #2}

\newcommand{\bool} {\ccode{bool}}
\newcommand{\dist}[1] {\ccode{dist} ~ #1}
\newcommand{\bind }{\vartriangleright  }

\newcommand{\comp}[1] { \ccode{Comp} ~ #1 }
\newcommand{\prob}[1] { \Pr\left[#1\right]}

\newcommand{\reals} { \mathds{R} }
\newcommand{\nat}{ \mathds{N} }

\newcommand{\lname}[1]{{{\textsf{#1}}}\xspace}
\newcommand{\libname}{\lname{Ceramist}}
\newcommand{\ssr}{\lname{Ssreflect}}
\newcommand{\mathcomp}{\lname{MathComp}}
\newcommand{\alea}{\lname{ALEA}} \newcommand{\fcf}{\lname{FCF}}
\newcommand{\ifth}{\lname{infotheo}}
\newcommand{\polaris}{\lname{Polaris}}

\newcommand{\prhl}{\lname{PRHL}}

\newcommand{\AMQ}{\lname{AMQ}}
\newcommand{\AMQMAP}{\lname{AMQMap}}

\newcommand{\shortadd}{\leftarrow_{\text{add}}}
\newcommand{\shorthash}{\leftarrow_{\text{hash}}}
\newcommand{\phash}{p_{\text{hash}}}

\definecolor{shadecolor}{gray}{1.00}
\definecolor{darkgray}{gray}{0.30}
\definecolor{violet}{rgb}{0.56, 0.0, 1.0}
\definecolor{forestgreen}{rgb}{0.13, 0.55, 0.13}

\lstdefinelanguage{Coq} {
mathescape=true,						
texcl=false,
morekeywords=[1]{
  Add,
  All,
  Arguments,
  Axiom,
  Bind,
  Canonical,
  Check,
  Close,
  CoFixpoint,
  CoInductive,
  Coercion,
  Contextual,
  Corollary,
  Defined,
  Definition,
  Delimit,
  End,
  Example,
  Export,
  Fact,
  Fixpoint,
  Goal,
  Graph,
  Hint,
  Hypotheses,
  Hypothesis,
  Implicit,
  Implicits,
  Import,
  Inductive,
  Lemma,
  Let,
  Local,
  Locate,
  Ltac,
  Maximal
  Module,
  Morphism,
  Next,
  Notation,
  Obligation,
  Open,
  Parameter,
  Parameters,
  Prenex,
  Print,
  Printing,
  Program,
  Projections,
  Proof,
  Proposition,
  Qed,
  Record,
  Relation,
  Remark,
  Require,
  Reserved,
  Resolve,
  Rewrite,
  Save,
  Scope,
  Search,
  Section,
  Show,
  Strict,
  Structure,
  Tactic,
  Theorem,
  Unset,
  Variable,
  Variables,
  View,
  inside,
  outside
},
morekeywords=[2]{
  as,
  cofix,
  else,
  end,
  exists,
  exists2,
  fix,
  for,
  forall,
  fun,
  if,
  in,
  is,
  let,
  match,
  nosimpl,
  of,
  return,
  struct,
  then,
  vfun,
  with,
  do,
  ret
},
morekeywords=[3]{Type, Prop, Set, True, False},
morekeywords=[4]{
  after,
  apply,
  assert,
  auto,
  bool_congr,
  case,
  change,
  clear,
  compute,
  congr,
  cut,
  cutrewrite,
  destruct,
  elim,
  field,
  fold,
  generalize,
  have,
  heval, 
  hnf,
  induction,
  injection,
  intro,
  intros,
  intuition,
  inversion,
  left,
  loss,
  move,
  nat_congr,
  nat_norm,
  pattern,
  pose,
  refine,
  rename,
  replace,
  revert,
  rewrite,
  right,
  ring,
  set,
  simpl,
  split,
  suff,
  suffices,
  symmetry,
  transitivity,
  trivial,
  under,
  unfold,
  unlock,
  using,
  without,
  wlog,
  autorewrite
},        
morekeywords=[5]{
  assumption,
  by,
  contradiction,
  done,
  exact,
  lia,
  gappa,
  omega,
  reflexivity,
  romega,
  solve,
  tauto,
  discriminate,
  unsat
},
morecomment=[s]{(*}{*)},
morekeywords=[6]{first, try, idtac, repeat},
showstringspaces=false,
morestring=[b]",
tabsize=3,							
extendedchars=true,  		 		
sensitive=true, 
breaklines=false,
basicstyle=\footnotesize\ttfamily,
captionpos=b,							
columns=[l]fullflexible,
identifierstyle={\color{black}},
keywordstyle=[1]{\color{violet}},
keywordstyle=[2]{\color{forestgreen}},
keywordstyle=[3]{\color{forestgreen}},
keywordstyle=[4]{\color{blue}},
keywordstyle=[5]{\color{red}},
keywordstyle=[6]{\color{violet}},
stringstyle=,
commentstyle=\it\ttfamily\color{brown},
numberstyle=\tiny,
literate={\\/}{{$\vee$}}1
         {/\\}{{$\wedge$}}1
         {:->}{{$\mapsto~$\!}}1
         {\\->}{{$\mapsto~$\!}}1
         {<--}{{$\asgn~$}}1
         {<-$}{{<-$\$~$}}1
         {\\in}{{$\in~$}}1
         {++}{{$+\!+\!~$}}1
         {->}{{$\to~$}}1
         {forall}{{$\forall~$}}1
         {exists}{{$\exists~$}}1
         {=>}{{$\Rightarrow~$}}1
         {\\+}{{$\!\join\!~$}}1
}

\lstdefinestyle{Coq}{language=Coq}

\let\subparagraph\paragraph

\usepackage{titlesec}
\titlespacing*{\section}{0pt}{*1.5}{*1}
\titlespacing*{\subsection}{0pt}{*1.5}{*1}
\titlespacing*{\paragraph}{0pt}{*1.0}{*1.0}
\setlist[itemize]{leftmargin=*}
\setlist[enumerate]{leftmargin=*}

\hypersetup{colorlinks,
  linkcolor=ACMDarkBlue,
  citecolor=ACMPurple,
  urlcolor=ACMDarkBlue,
  filecolor=ACMDarkBlue}

\begin{document}


\title{Certifying Certainty and Uncertainty\\ 
  in Approximate Membership Query Structures}

\titlerunning{Certifying Certainty and Uncertainty in AMQs}
\authorrunning{}


\vspace{-5pt}

\author{
  Kiran Gopinathan\inst{1} \and
  Ilya Sergey\inst{2,1}
}

\institute{
  School of Computing, National University of Singapore, Singapore
  \and
  Yale-NUS College, Singapore
  \\
  \email{\{kirang,ilya\}@comp.nus.edu.sg}
}




\maketitle

\vspace{-10pt}

\begin{abstract}
Approximate Membership Query structures (AMQs) rely on randomisation
for time- and space-efficiency, while introducing a possibility of
false positive and false negative answers.
Correctness proofs of such structures involve subtle reasoning about
bounds on probabilities of getting certain outcomes.
%
%
Because of these subtleties, a number of unsound arguments in such
proofs have been made over the years.

In this work, we address the challenge of building rigorous and
reusable computer-assisted proofs about probabilistic specifications of
AMQs.
We describe the framework for systematic decomposition of AMQs and
their properties into a series of interfaces and reusable components.
We implement our framework as a library in the Coq proof assistant and
showcase it by encoding in it a number of non-trivial AMQs, such as
Bloom filters, counting filters, quotient filters and blocked
constructions, and mechanising the proofs of their probabilistic
specifications.

We demonstrate how AMQs encoded in our framework guarantee the absence
of false negatives \emph{by construction}.
We also show how the proofs about probabilities of false positives for
complex AMQs can be obtained by means of \emph{verified reduction} to
the implementations of their simpler counterparts.
Finally, we provide a library of domain-specific theorems and tactics
that allow a high degree of automation in probabilistic proofs.




\end{abstract}

\vspace{-15pt}

\section{Introduction}
\label{sec:introduction}

Approximate Membership Query structures (AMQs) are probabilistic data
structures that compactly implement (multi-)sets via hashing.
%
%
They are a popular alternative to traditional collections in
algorithms whose utility is not affected by some fraction of wrong
answers to membership queries.
Typical examples of such data structures are Bloom
filters~\cite{Bloom:1970:STH:362686.362692}, quotient
filters~\cite{BenderFJKKMMSSZ12,PaghPR05}, and count-min
sketches~\cite{CormodeM05}.
In particular, versions of Bloom filters find many applications in
security and privacy~\cite{ErlingssonPK14,GerbetKL15,NaorY19}, static
program analysis~\cite{NasreRGK09}, databases~\cite{cassandra-bloom},
web search~\cite{GoodwinHLCCEH17}, suggestion
systems~\cite{medium-bloom}, and blockchain
protocols~\cite{eth-bloom,GervaisCKG14}.

Hashing-based AMQs achieve efficiency by means of losing precision
when answering queries about membership of certain elements. Luckily,
most of the applications listed above can tolerate \emph{some} loss of
precision.
For instance, a static points-to analysis may consider two memory
locations as aliases even if they are not (a \emph{false positive}),
still remaining sound. However, it would be unsound for such an
analysis to claim that two locations do not alias in the case they
do (a \emph{false negative}).
%
%
Even if it increases the number of false positives, a randomised data
structure can be used to answer aliasing queries in a sound way---as
long as it does not have false negatives~\cite{NasreRGK09}.
But \emph{how much} precision would be lost if, \eg, a Bloom filter
with certain parameters is chosen to answer these queries?
Another example, in which quantitative properties of false positives
are critical, is the security of Bitcoin’s Nakamoto
consensus~\cite{Nakamoto:08} that depends on the counts of block
production per unit time~\cite{GervaisCKG14}.

In the light of the described above applications, of particular
interest are two kinds of properties specifying the behaviour of AMQs:
\begin{itemize}
\item \emph{No-False-Negatives} properties, stating that a
  set-membership query for an element $x$ always returns \lname{true}
  if $x$ is, in fact, in the set represented by the~AMQ.

\item Properties quantifying the rate of \emph{False Positives} by
  providing a probabilistic bound on getting a wrong ``yes''-answer
  to a membership query, given certain parameters of the data
  structure and the past history of its usage.
\end{itemize}
Given the importance of such claims for practical applications, it is
desirable to have machine-checked formal proofs of their validity.
And, since many of the existing AMQs share a common design structure,
one may expect that a large portion of those validity proofs can be
reused across different implementations.

Computer-assisted reasoning about the absence of \emph{false
  negatives} in a particular AMQ (Bloom filter) has been addressed to
some extent in the past~\cite{BlotDL16}.
%
%
However, to the best of our knowledge, mechanised proofs of
probabilistic bounds on the \emph{rates of false positives} did
not extend to such structures.
Furthermore, to the best of our knowledge, no other existing AMQs have
been formally verified to date, and no attempts were made towards
characterising the commonalities in their implementations in order to
allow efficient proof reuse.

%

In this work, we aim to advance the state of the art in
machine-checked proofs of probabilistic theorems about false positives
in randomised hash-based data structures. As recent history
demonstrates, when done in a ``paper-and-pencil'' way, such proofs may
contain subtle mistakes~\cite{Bose2008Oct,christensen2010new} due to
misinterpreted assumptions about relations between certain kinds of
events.
These mistakes are not surprising, as the proofs often need to perform
a number complicated manipulations with expressions that capture
probabilities of certain events.
Our goal is to factor out these reasoning patterns into a standalone
library of \emph{reusable} program- and specification-level
definitions and theorems, implemented in a proof assistant enabling
computer-aided verification of a variety of AMQs.


\paragraph{Our contributions.~}

The key novel observation we make in this work is the decomposition of
the common AMQ implementations into the following components: (a) a
hashing strategy and (b) a state component that operates over hash
outcomes,
together capturing most AMQs that provide fixed constant-time
insertion and query operations.
%
%
%
Any AMQ that is implemented as an instance of those components enjoys
the \emph{no-false-negatives} property \emph{by construction}.
Furthermore, such a decomposition streamlines the proofs of
structure-specific bounds on false positive rates, while allowing
for proof reuse for complex AMQ implementations, which are built on
top of simpler AMQs~\cite{PutzeSS09}.
Powered by those insights, this work makes the following technical
contributions:

\begin{itemize}

\item A Coq-based mechanised framework \libname,
  specialised for reasoning about AMQs.\footnote{\libname stands for
    \textbf{Cer}tified \textbf{A}pproximate
    \textbf{M}embersh\textbf{i}p \textbf{St}ructures.}
  Implemented as a Coq library, it provides a systematic decomposition
  of AMQs and their properties in terms of Coq modules and uses these
  interfaces to to derive certain properties ``for free'', as well as
  supporting proof-by-reduction arguments between classes of similar
  AMQs.

\item A library of non-trivial theorems for expressing closed-form
  probabilities on false positive rates in AMQs. In particular, we
  provide the first mechanised proof of the closed form for Stirling
  numbers of the second kind~\cite[Chapter~6]{GKP1994}.

\item A collection of proven facts and tactics for effective
  construction of proofs of probabilistic properties. Our approach
  adopts the style of \ssr
  reasoning~\cite{Gonthier-al:TR,Maboubi-Tassi:MathComp}, and
  expresses its core lemmas in terms of rewrites and evaluation.

\item A number of case study AMQs mechanised via \libname:
  ordinary~\cite{Bloom:1970:STH:362686.362692} and
  counting~\cite{TarkomaRL12} Bloom filters, quotient
  filters~\cite{BenderFJKKMMSSZ12,PaghPR05}, and Blocked
  AMQs~\cite{PutzeSS09}.

\end{itemize}

\noindent
For ordinary Bloom filters, we provide the first mechanised proof that
the probability of a false positive in a Bloom filter can be written
as a closed form expression in terms of the input parameters; a bound
that has often been mis-characterised in the past due to oversight of
subtle dependencies between the components of the
structure~\cite{Bloom:1970:STH:362686.362692,mitzenmacher2005}.
For Counting Bloom filters, we provide the first mechanised proofs of
several of their properties: that they have no false negatives, its false
positive rate, that an element can be removed without affecting
queries for other elements, and the fact that Counting Bloom filters
preserve the number of inserted elements irrespective of the
randomness of the hash outputs.
  %
%
For quotient filters, we provide a mechanised proof of the false
positive rate and of the absence of false negatives.
Finally, alongside the standard Blocked Bloom filter~\cite{PutzeSS09},
we derive two novel AMQ data structures: \emph{Counting Blocked Bloom
  filters} and \emph{Blocked Quotient filters}, and prove
corresponding no-false-negatives and false positive rates for all of
them.
Our case studies illustrate that \libname can be repurposed to verify
hash-based AMQ structures, including entirely new ones that have not
been described in the literature, but rather have been obtained by
composing existing AMQs via the ``blocked'' construction.

Our mechanised development~\cite{ceramist} is entirely
\emph{axiom-free}, and is compatible with Coq~8.11.0~\cite{Coq-manual}
and \mathcomp~1.10~\cite{Maboubi-Tassi:MathComp}.
It relies on the \ifth library~\cite{AffeldtHS14} for encoding
discrete probabilities.
%

\paragraph{Paper~outline.~}

We start by providing the intuition on Bloom filters, our main
motivating example, in \autoref{sec:overview}.
We proceed by explaining the encoding of their semantics, auxiliary
hash-based structures, and key properties in Coq in
\autoref{sec:encoding}.
\autoref{sec:framework} generalises that encoding to a general AMQ
interface, and provides an overview of \libname, its embedding into
Coq, showcasing it by another example instance---Counting Bloom
filters.
\autoref{sec:bigsum} describes the specific techniques that help to
structure our mechanised proofs.
In \autoref{sec:casestudies}, we report on the evaluation of \libname
on various case studies, explaining in detail our compositional
treatment of blocked AMQs and their properties.
\autoref{sec:related} provides a discussion on the state of the
art in reasoning about probabilistic data structures.






\section{Motivating Example}
\label{sec:overview}


\libname is a library specialised for reasoning about AMQ data
structures in which the underlying randomness arises from the
interaction of one or more hashing operations.
To motivate this development, we thus consider applying it to the
classical example of such an algorithm---a Bloom
filter~\cite{Bloom:1970:STH:362686.362692}.
%

\subsection{The Basics of Bloom Filters}
\label{sec:whtisblmfilter}


Bloom filters are probabilistic data structures that provide compact
encodings of mathematical sets, trading increased space efficiency for
a weaker membership test~\cite{Bloom:1970:STH:362686.362692}.
Specifically, when testing membership for a value \emph{not} in the
Bloom filter, there is a possibility that the query may be answered as
positive.
Thus a property of direct practical importance is the exact
probability of this event, and how it is influenced by the other
parameters of the implementation.
%


%
\begin{wrapfigure}[10]{r}{0.45\textwidth}
\vspace{-22pt}
\includegraphics[width=0.45\textwidth]{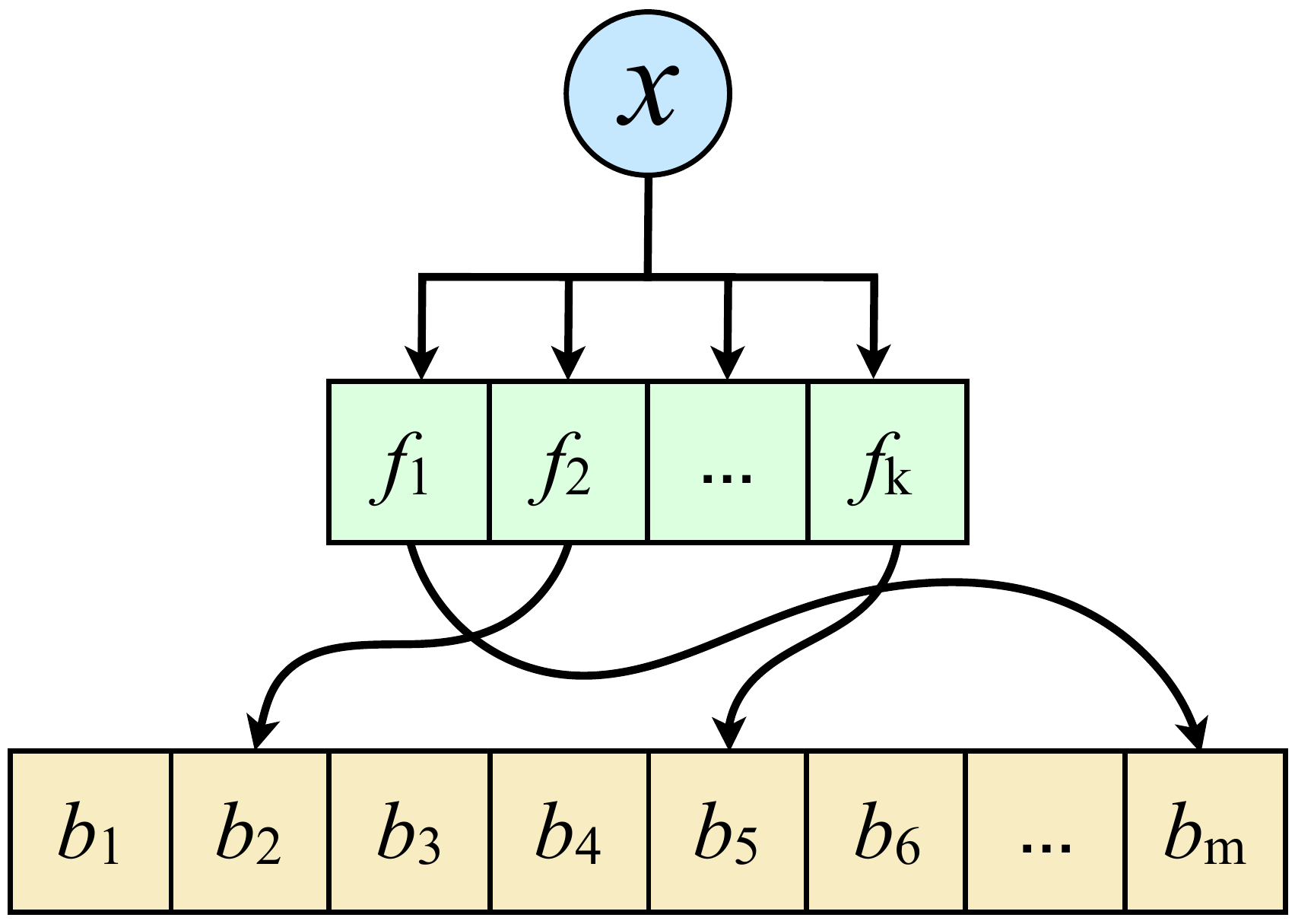}
\end{wrapfigure}
A Bloom filter $\bloomfilter$ is implemented as a binary vector of $m$
bits (all initially zeros), paired with a sequence of $k$ hash
functions $f_1, \dots, f_k$, 
collectively mapping each input value to a vector of $k$ indices from
$\set{1 \ldots m}$, the indices determine the bits set to
\textsf{true} in the $m$-bit array
%
%
Assuming an ideal selection of hash functions, we can treat the output
of $f_1, \dots, f_k$ on new values as a uniformly-drawn random vector.
To insert a value $x$ into the Bloom filter, we can treat each element
of the ``hash vector'' produced from $f_1,\dots,f_k$ as an index into
$\bloomfilter$ and set the corresponding bits to ones.
Similarly, to test membership for an element $x$, we can check that
all $k$ bits specified by the hash-vector are raised.

\subsection{Properties of Bloom Filters}
\label{sec:properties}

Given this model, there are two obvious properties of practical
importance: that of false positives and of false negatives.

\paragraph{False Negatives.}

It turns out that these definitions are sufficient to guarantee the
lack of false-negatives with complete certainty, \ie, irrespective of
the random outcome of the hash functions. This follows from the fact
that once a bit is raised, there are no permitted operations that will
unset it.

\begin{theorem}[No False Negatives]
\label{lemma:overview-fneg}
If~~$x~\in~\bloomfilter$, then
$\prob{\bfquery{x}{\bloomfilter}} = 1$,
where $\bfquery{x}{\bloomfilter}$ stands for the approximate
membership test, while the relation $x \in \bloomfilter$ means that
$x$ has been previously inserted into $\bloomfilter$.
\end{theorem}

%

%
\paragraph{False Positives.}

This property is more complex as the occurrence of a false positive is
entirely dependent on the particular outcomes of the hash functions
$f_1,\dots,f_k$ and one needs to consider situations in which the hash
functions happen to map some values to \emph{overlapping} sets of
indices.
That is, after inserting a series of values $\xseq$, subsequent
queries for $y \notin \xseq$ might incorrectly return \lname{true}.

This leads to subtle dependencies that can invalidate the analysis,
and have lead to a number of incorrect probabilistic bounds on the
event, including in the analysis by Bloom in his original
paper~\cite{Bloom:1970:STH:362686.362692}.
Specifically, Bloom first considered the probability that inserting
$l$ distinct items into the Bloom filter will set a particular bit
$b_i$. From the independence of the hash functions, he was able to
show that the probability of this event has a simple closed-form
representation:
\begin{lemma}[Probability of a single bit being set]
\label{lemma:overview-singbit}
If the only values previously inserted into $\bloomfilter$ are
$x_1,\dots,x_l$, then the probability of a particular single bit at
the position $i$ being set is
$
    \prob{\text{\emph{$i^{\text{th}}$ bit in $\bloomfilter$ is set}}} =
    1 - \left( 1 - \frac{1}{m}\right)^{kl}.
$
\end{lemma}
Bloom then claimed that the probability of a false positive was simply
the probability of a single bit being set, raised to the power of $k$,
reasoning that a false positive for an element
$y \not\in \bloomfilter$ only occurs when all the $k$ bits
corresponding to the hash outputs are set.

Unfortunately, as was later pointed out by Bose
\etal~\cite{Bose2008Oct}, as the bits specified by $f_1(x),\dots,f_{k-1}(x)$
may overlap, we cannot guarantee the independence that is required for
any simple relation between the probabilities.
Bose \etal rectified the analysis by instead interpreting the bits
within a Bloom filter as maintaining a set
$\text{bits}(\bloomfilter) \subseteq \nat_{[0,\dots,m-1]}$,
corresponding to the indices of raised bits.
With this interpretation, an element $y$ only tests positive if the
random set of indices produced by the hash functions on $y$ is such
that $\text{inds}(y) \subseteq \text{bits}(\bloomfilter)$.
Therefore, the chance of a positive result for
$y \not\in \bloomfilter$ resolves to the chance that the random set of
indices from hashing $y$ is a subset of the union of $\text{inds}(x)$
for each $x \in \bloomfilter$.
The probability of this reduced event is described by the following
theorem:

\begin{theorem}[Probability of False Positives]
  If the only values inserted into $\bloomfilter$ are $x_1,\dots,x_l$,
  then for any $y \not\in \bloomfilter$,
$
    \prob{\bfquery{y}{\bloomfilter}} = \frac{1}{m^{k(l + 1)}} \sum_{i
      = 1}^{m} i^k i! \ncr{m}{i} \stirlingsnd{kl}{i},
$
where $\stirlingsnd{s}{t}$ stands for the \emph{Stirling number of the
  second kind}, capturing the number of surjections from a set of size
$s$ to a set of size $t$.
\end{theorem}

The key step in capturing these program properties is in treating the
outcomes of hashes as \emph{random variables} and then propagating
this randomness to the results of the other operations.
A formal treatment of program outcomes requires a suitable semantics,
representing programs as distributions of such random variables.
In moving to mechanised proofs, we must first fully characterise this
semantics, formally defining a notion of a probabilistic computation
in Coq.




\section{Encoding AMQs in Coq}
\label{sec:encoding}

To introduce our encoding of AMQs and their probabilistic behaviours
in Coq, we continue with our running example, transitioning from
mathematical notation to Gallina, Coq's language.
The rest of this section will introduce each of the key components of
this encoding through the lens of Bloom filters.

\subsection{Probability Monad}
\label{sec:prob-monad}


Our formalisation represents probabilistic computations using an
embedding following the style of the \fcf
library~\cite{Petcher-Morissett:POST15}.
We do not use \fcf directly, due to its primary focus on cryptographic
proofs, wherein it provides little support for proving probabilistic
bounds directly, instead prioritising a reduction-based approach of
expressing arbitrary computations as compositions of known
distributions.

Following the adopted \fcf notation, a term of type $\comp{A}$
represents a probabilistic computation returning a value of type $A$,
and is constructed using the standard monadic operators, with an
additional primitive $\ccode{rand} ~ n$ that allows sampling from a
uniform distribution over the range $\ordnat{n}$:
\begin{align*}
\ccode{ret} & : A \rightarrow \comp{ A } \\
\ccode{bind} & : \comp{A} \rightarrow (A \rightarrow \comp B ) \rightarrow \comp{ B } \\
\ccode{rand} & : (n: \nat) \rightarrow \comp{\left(\ordnat{n}\right)}
\end{align*}
We implement a Haskell-style \code{do}-notation over this monad to
allow descriptions of probabilistic computations within Gallina. For
example, the following code is used to implement the query operation
for the Bloom filter:
\begin{lstlisting}
  hash_res <-$ hash_vec_int x hashes; (* hash x using the hash functions *)
  let (new_hashes, hash_vec) := hash_res in
  (* check if all the corresponding bits are set *)
  let qres := bf_query_int hash_vec bf in
  (* return the query result and the new hashes *)
       ret (new_hashes, qres).
  \end{lstlisting}
%
%
%
  In the above listing, we pass the queried value $\ccode{x}$ along
  with the hash functions $\ccode{hashes}$ to a probabilistic
  hashing operation $\ccode{hash_vec_int}$ to hash $\ccode{x}$ over
  each function in $\ccode{hashes}$.
The result of this random operation is then bound to
$\ccode{hash_res}$
and split into its constituent components---a sequence of hash outputs
$\ccode{hash_vec}$ and an updated copy $\ccode{new_hashes}$ of the
hash functions, now incorporating the mapping for
\ccode{x}. 
%
Then, having mapped our input into a sequence of indices, we can query
the Bloom filter for membership using a corresponding deterministic
operation $\ccode{bf_query_int}$ to check that all the bits specified
by $\ccode{hash_vec}$ are set.
Finally, we complete the computation by returning the query outcome
$\ccode{qres}$ and the updated hash functions $\ccode{new_hashes}$
using the $\ccode{ret}$ operation to lift our result to a
probabilistic outcome.

Using the code snippet above, we can define the query operation
$\ccode{bf_query}$ as a function that maps a Bloom filter, a value to
query, and a collection of hash functions to a probabilistic
computation returning the query result and an updated set of hash
functions.
However, because our computation type does not impose any particular
semantics, this result only encodes the \emph{syntax} of the
probabilistic query and has no actual meaning without a separate
interpretation.


Thus, given a Gallina term of type $\comp{A}$, we must first evaluate
it into a distribution over possible results to state properties on
the probabilities of its outcomes.
We interpret our monadic encoding in terms of Ramsey's probability
monad~\cite{Ramsey-Pfeffer:POPL02}, which decomposes a complex
distribution into composition of primitive ones bound together via
conditional distributions.
%
%
To capture this interpretation within Coq, we then use the encoding of
this monad from the \ifth
library~\cite{Affeldt-Hagiwara:ITP12,AffeldtHS14}, and provide a
function $\ccode{eval_dist} : \comp A \rightarrow \dist A$ that
evaluates computations into distributions by recursively mapping them
to the probability monad.
%
%
%
Here, $\ccode{dist A}$ represents \ifth's encoding of distributions
over a finite support $\ccode{A}$, defined as being composed of a
measure function $\ccode{pmf}: A \rightarrow \reals^+$, and a proof
that the sum of the measure over the support $A$ produces~1.







This mapping from computations to distributions must be done to a
program $e$ (involving, \eg, Bloom filter) before stating its
probability bound.
%
%
Therefore, we hide this evaluation process behind a notation that
allows stating probabilistic properties in a form closer to their
mathematical counterparts:
%
%
\begin{align*}
  \prob{ e = v }  & \eqdef (\ccode{eval_dist} ~  e) ~ v  \\
  \prob{ e }  & \eqdef (\ccode{eval_dist} ~  e) ~ \mathsf{true}  \\
\end{align*}
Above, $v$ is an arbitrary element in the support of the distribution
induced by $e$.
Finally, we introduce a binding operator $\bind$ to allow concise
representation of dependent distributions:
$e \bind f \eqdef \ccode{bind} ~ e ~ f$.


\subsection{Representing Properties of Bloom Filters}
\label{sec:props}

We define the state of a Bloom filter (\ccode{BF}) in Coq as a binary
vector of a fixed length $m$, using \ssr's $\ccode{m.-tuple}$ data
type:
%
%
\begin{lstlisting}
Record BF := mkBF { bloomfilter_state: m.-tuple bool }.
Definition bf_new : BF :=  (* construct a BF with all bits cleared *).
Definition bf_get_int i : BF -> bool :=   (* retrieve BF's ith bit *).
\end{lstlisting}
%
We define the deterministic components of the Bloom filter
implementation as pure functions taking an instance of \ccode{BF} and
a series of indices assumed to be obtained from earlier calls to the
associated hash functions:

\vspace{-5pt}                   %

{\footnotesize{
\begin{align*}
\ccode{bf_add_int} & : \ccode{BF} \rightarrow \ccode{seq} ~  \ordnat{m}  \rightarrow \ccode{BF} \\
\ccode{bf_query_int} & : \ccode{BF} \rightarrow \ccode{seq} ~  \ordnat{m}  \rightarrow \bool
\end{align*}
}}
That is, $\ccode{bf_add_int}$ takes the Bloom filter state and a
sequence of indices to insert and returns a new state with the
requested bits also set.
Conversely, $\ccode{bf_query_int}$ returns \lname{true} \Iff all the
queried indices are set.
These pure operations are then called within a probabilistic wrapper
that handles hashing the input and the book-keeping associated with
hashing to provide the standard interface for AMQs:


\vspace{-10pt}

{\footnotesize{
\begin{align*}
  \ccode{bf_add} & :  B \rightarrow (\ccode{HashVec} ~ B * \ccode{BF}) \rightarrow \comp{(\ccode{HashVec} ~ B * \ccode{BF})} \\
  \ccode{bf_query}
                 & :  B \rightarrow (\ccode{HashVec} ~ B * \ccode{BF}) \rightarrow \comp{(\ccode{HashVec} ~ B * \bool)}
\end{align*}
}}

\vspace{-10pt}
\noindent
The component $\ccode{HashVec} ~ B$ (to be defined in
\autoref{sec:hashvecprim}), parameterised over an input type $B$,
keeps track of \emph{known results} of the involved hash functions and
is provided as an external parameter to the function rather than being
a part of the data structure to reflect typical uses of AMQs, wherein
the hash operation is pre-determined and shared by \emph{all}
instances.


With these definitions and notation, we can now state the main
theorems of interest about Bloom filters directly within
Coq:\footnote{\texttt{bf\_addm} is a trivial generalisation of the
  insertion to multiple elements.}
%


\begin{theorem}[No False Negatives] For any Bloom filter state $\bloomfilter$, a vector of
  hash functions $\hshs$, after having inserted an element $x$ into
  $\bloomfilter$, followed by a series $\xseq$ of other inserted
  elements, the result of query $\bfquery{x}{\bloomfilter}$ is always
  \emph{\lname{true}}.
  That is, in terms of probabilities: 
$  \prob{\cccode{bf_add} ~ x ~ (\hshs,\bloomfilter) \bind
  \cccode{bf_addm} ~ \xseq \bind \cccode{bf_query} ~ x }~=~1.
$
\label{lemma:enc-fneg}
\end{theorem}

\vspace{-15pt}

\begin{lemma}[Probability of Flipping a Single Bit]
\label{thm:sing-bit}
For a vector of hash functions $\hshs$ of length $k$, after inserting
a series of $l$ distinct values $\xseq$, all unseen in $\hshs$, into
an empty Bloom filter $\bloomfilter$, represented by a vector of $m$
bits, the probability of its any index $i$ being set is
$ \prob{\cccode{bf_addm} ~ \xseq ~ (\hshs,\cccode{bf_new}) \bind
  \cccode{bf_get} ~ i } = 1 - \left( 1 - \frac{1}{m}\right)^{kl}. $
Here, \cccode{bf_get} is a simple embedding of the pure function
\cccode{bf_get_int} into a probabilistic computation.
\end{lemma}


\begin{theorem}[Probability of a False Positive] 
  After having inserted a series of $l$ distinct values $\xseq$, all
  unseen in $\hshs$, into an empty Bloom filter $\bloomfilter$, for
  any unseen $y \not\in \xseq$, the probability of a subsequent query
  $\bfquery{y}{\bloomfilter}$ for $y$ returning \lname{true} is given
  as
  $
    \prob{  \cccode{bf_addm} ~ \xseq ~ (\hshs, \cccode{bf_new})  \bind  \cccode{bf_query} ~ y }
    = 
    \frac{1}{m^{k(l + 1)}} \sum_{i = 1}^{m} i^k i! \ncr{m}{i} \stirlingsnd{kl}{i}.
  $  
\label{lemma:enc-fpos}
\end{theorem}

\noindent
The proof of this theorem required us to provide \emph{the first
  axiom-free mechanised proof} for the closed form for Stirling
numbers of the second kind~\cite{GKP1994}.
%

In the definitions above, we used the output of the hashing operation
as the bound between the deterministic and probabilistic components
of the Bloom filter.
For instance, in our earlier description of the Bloom filter query
operation in \autoref{sec:prob-monad}, we were able to implement
the entire operation with the only probabilistic operation being the
call \ccode{hash_vec_int x hashes}.
In general, structuring AMQ operations as manipulations with hash
outputs via \emph{pure} deterministic functions allows us to decompose
reasoning about the data structure into a series of specialised
properties about its deterministic primitives and a separate set of
reusable properties on its hash operations.



\subsection{Reasoning about Hash Operations} %
\label{sec:hashvecprim}

%
%
%
%
%

We encode hash operations within our development using a random
oracle-based implementation.
In particular, in order to keep track of \emph{seen} hashes learnt by
hashing previously observed values, we represent a \emph{state} of a
hash function from elements of type \ccode{B} to a range $\ordnat{m}$
using a finite map to ensure that previously hashed values produce the
same hash output:
\begin{lstlisting}
Definition HashState B := FixedMap B 'I_m.
\end{lstlisting}
The state is paired with a hash function generating uniformly random
outputs for unseen values, and otherwise returns the value as from its
prior invocations:
\begin{lstlisting}
Definition hash value state : Comp (HashState B * B) :=
  match find value state  with
  | Some(output) => ret (state, output)
  | None => rnd <-$ rand m;
             new_state <- put value rnd state;
             ret (new_state, rnd)
  end.
\end{lstlisting}
%
A \emph{hash vector} is a generalisation of this structure to
represent a vector of states of $k$ independent hash functions:
\begin{lstlisting}
Definition HashVec B := k.-tuple HashState B.
\end{lstlisting}
The corresponding hash operation over the hash vector,
$\ccode{hash_vec_int}$, is then defined as a function taking a value
and the current hash vector and then returning a pair of the updated
hash vector and associated random vector, internally calling out to
$\ccode{hash}$ to compute individual hash outputs.






This random oracle-based implementation allows us to formulate several
helper theorems for simplifying probabilistic computations using
hashes by considering whether the hashed values \emph{have been seen
  before or not}.
For example, if we knew that a value $x$ had not been seen before, we
would know that the possibility of obtaining any particular choice of
a vector of indices would be equivalent to obtaining the same vector
by a draw from a corresponding uniform distribution. We can formalise
this intuition in the form of the following theorem:


\begin{theorem}[Uniform Hash Output]
\label{thm:hash-uniform}
For any two hash vectors $\hshs$, $\hshs'$ of length $k$, a value $x$
that has not been hashed before, and an output vector $\inds$ of
length $m$ obtained by hashing $x$ via $\hshs$,
if the state of $\hshs'$ has the same mappings as $\hshs$ and also
maps $x$ to $\inds$, the probability of obtaining the pair
$(\hshs',\inds)$ is uniform:
$\prob{\cccode{hash_vec_int} ~ x ~ \hshs = (\hshs',\inds)} =
\left(\frac{1}{m}\right)^{k}$
\end{theorem}

\noindent
Similarly, there are also often cases where we are hashing a value
that we \emph{have already seen}. In these cases, if we know the exact
indices a value hashes to, we can prove a certainty on the value of
the outcome:
\begin{theorem}[Hash Consistency]
\label{thm:reducing-hash-vector}
  For any hash vector $\hshs$, a value $x$,
  if $\hshs$ maps $x$ to outputs $\inds$, then hashing $x$ again will
  certainly produce $\inds$ and not change $\hshs$, that is,
  $\prob{\cccode{hash_vec_int} ~ x ~ \hshs = (\hshs,\inds)} = 1$.
\end{theorem}

\noindent
By combining these types of probabilistic properties about hashes with
the earlier Bloom filter operations, we are able to prove the prior
theorems about Bloom filters by reasoning primarily about the core
logical interactions of the \emph{deterministic components} of the
data structure.
This decomposition is not just applicable to the case of Bloom
filters, but can be extended into a general framework for obtaining
modular proofs of AMQs, as we will show in the next section.

%


\section{\libname at Large}
\label{sec:framework}

Zooming out from the previous discussion of Bloom filters, we now
present \libname in its full generality, describing the high-level
design in terms of the various interfaces it requires to instantiate
to obtain verified AMQ implementations.

The core of our framework revolves around the decomposition of an AMQ
data structure into separate interfaces for hashing (\lname{AMQHash})
and state (AMQ), generalising the specific decomposition used for
Bloom filters (hash vectors and bit vectors respectively). More
specifically, the \lname{AMQHash} interface captures the probabilistic
properties of the hashing operation, while the \AMQ interface captures
the deterministic interactions of the state with the hash outcomes.

\subsection{\lname{AMQHash} Interface}
\label{sec:amqhash-interface}

The \lname{AMQHash} interface generalises the behaviours of hash vectors
(\autoref{sec:hashvecprim}) to provide a generic description of the
hashing operation used in AMQs.

The interface first abstracts over the specific types used in the
prior hashing operations (such as, \eg, \ccode{HashVec B}) by treating
them as opaque parameters:
using a parameter $\ccode{AMQHashState}$ to represent the state of the
hash operation; types $\ccode{Key}$ and $\ccode{Value}$ encoding the
hash inputs and outputs respectively, and finally, a deterministic operation
$\ccode{AMQHash_add_internal} : \ccode{AMQHashState} \rightarrow \ccode{Key} \rightarrow \ccode{Value} \rightarrow \ccode{AMQHashState}$ to
encode the interaction of the state with the outputs and inputs.
%
%
For example, in the case of a single hash, the state parameter
\ccode{AMQHashState} would be \ccode{HashState B}, while for a hash
vector this would instead be \ccode{HashVec B}.

To use this hash state in probabilistic computations, the interface
assumes a separate probabilistic operation that will take the hash
state and randomly generate an output (\eg, \lstinline{hash} for
single hashes and \lstinline{hash_vec_int} for hash vectors):
\begin{lstlisting}
Parameter AMQHash_hash: Key -> AMQHashState -> Comp (AMQHash * Value).
\end{lstlisting}

Then, to abstractly capture the kinds of reasoning about the outcomes
of hash operations done with Bloom filters in
\autoref{sec:hashvecprim}, the interface assumes a few predicates
on the hash state to provide information about its contents:

\begin{lstlisting}
Parameter AMQHash_hashstate_contains: AMQHashState -> Key -> Value -> bool.
Parameter AMQHash_hashstate_unseen: AMQHashState -> Key -> bool.
\end{lstlisting}

These components are then combined together to produce more abstract
formulations of the previous Theorems~\ref{thm:hash-uniform} and
\ref{thm:reducing-hash-vector} on hash operations.
%

\vspace{3pt}

\begin{prop}[Generalised Uniform Hash Output]
\label{thm:gen-hash-uniform}
  There exists a probability $\phash$, such that for any two AMQ
  hash states $\hshs,\hshs'$, a value $x$ that is unseen, and an
  output $\inds$ obtained by hashing $x$ via $\hshs$,
  if the state of $\hshs'$ has the same mappings as $\hshs$ and also
  maps $x$ to $\inds$, the probability of obtaining the pair
  $(\hshs',\inds)$ is given by:
  $\prob{\cccode{AMQHash\_hash} ~ x ~ \hshs = (\hshs',\inds)} = \phash$.
\label{thm:amqhash-generalised-uniform}
\end{prop}

\vspace{3pt}

\begin{prop}[Generalised Hash Consistency]
\label{thm:general-reducing-hash-vector}
  For any AMQ hash state $\hshs$, a value $x$,
  if $\hshs$ maps $x$ to an output $\inds$, then hashing $x$ again
  will certainly produce $\inds$ and not change $\hshs$:
  $\prob{\cccode{AMQhash\_hash} ~ x ~ \hshs = (\hshs,\inds)} = 1$
\end{prop}

\vspace{3pt}

Proofs of these corresponding properties must also be provided to
instantiate the \lname{AMQHash} interface.
Conversely, components operating over this interface can assume their
existence, and use them to abstractly perform the same kinds of
simplifications as done with Bloom filters, resolving many
probabilistic proofs to dealing with deterministic properties
on the AMQ states.

\subsection{The \AMQ Interface}
\label{sec:amq-interface}

Building on top of an abstract \lname{AMQHash} component, the \AMQ
interface then provides a unified view of the state of an AMQ and how
it deterministically interacts with the output type \ccode{Value} of a
particular hashing operation.

As before, the interface begins by abstracting the specific types and
operations of the previous analysis of Bloom filters, first
introducing a type $\ccode{AMQState}$ to capture the state of the AMQ,
and then assuming deterministic implementations of the typical
\emph{add} and \emph{query} operations of an AMQ:
\begin{lstlisting}
Parameter AMQ_add_internal: AMQState -> Value -> AMQState.
Parameter AMQ_query_internal: AMQState -> Value -> bool.
\end{lstlisting}
In the case of Bloom filters, these would be instantiated with the
\lstinline{BF}, \lstinline{bf_add_int} and \lstinline{bf_query_int}
operations respectively (\cf~\autoref{sec:props}), thereby setting the
associated hashing operation to the hash vector
(\autoref{sec:hashvecprim}).

As we move on to reason about the behaviours of these operations, the
interface diverges slightly from that of the Bloom filter by
conditioning the behaviours on the assumption that the state has
sufficient capacity:
\begin{lstlisting}
Parameter AMQ_available_capacity: AMQState -> nat -> bool.
\end{lstlisting}
While the Bloom filter has no real deterministic notion of a capacity,
this cannot be said of all AMQs in general, such as the Counting Bloom
filter or Quotient filter, as we will discuss later.

With these definitions in hand, the behaviours of the AMQ operations
are characterised using a series of associated assumptions:

\vspace{3pt}

\begin{prop}[AMQ insertion validity]
  \label{thm:amq-insert-validity} For a state $s$ with sufficient
  capacity, inserting any hash output $\inds$ into $s$ via
  \cccode{AMQ_add_internal} will produce a new state $s'$ for which
  any subsequent queries for $\inds$ via $\cccode{AMQ_query_internal}$
  will return \emph{\lname{true}}.
\end{prop}

\vspace{3pt}

\begin{prop}[AMQ query preservation]
\label{thm:amq-query-pres}
For any AMQ state $s$ with sufficient remaining capacity, if queries
for a particular hash output $\inds$ in $s$ via
\cccode{AMQ_query_internal} happen to return \emph{\lname{true}}, then
inserting any further outputs $\inds'$ into $s$ will return a state
for which queries for $\inds$ will \emph{still} return
\emph{\lname{true}}.
\end{prop}

\vspace{3pt}

Even though these assumptions seemingly place strict restrictions on
the permitted operations, we found that these properties are satisfied
by most common AMQ structures.
One potential reason for this might be because they are in fact
\emph{sufficient} to ensure the No-False-Negatives property standard
of most AMQs:
\begin{theorem}[Generalised No False Negatives] 
  For any AMQ state $s$, a corresponding hash state $\hshs$, after
  having inserted an element $x$ into $s$, followed by a series
  $\xseq$ of other inserted elements, the result of query for $x$ is
  always \emph{\lname{true}}.
  That is,
$  \prob{\cccode{AMQ_add} ~ x ~ (\hshs,s) ~\bind  ~ \cccode{AMQ_addm}
  ~ \xseq~\bind ~ \cccode{AMQ_query} ~ x } = 1.
$
\label{lemma:generalised-enc-fneg}
\end{theorem}

\noindent
Here, \ccode{AMQ_add}, \ccode{AMQ_addm}, and \ccode{AMQ_query} are
generalisations of the probabilistic wrappers of Bloom filters
(\cf~\autoref{sec:prob-monad}) for doing the bookkeeping associated
with hashing and delegating to the internal deterministic operations.

The generalised \autoref{lemma:generalised-enc-fneg} illustrates one
of the key facilities of our framework, wherein by simply providing
components satisfying the \lname{AMQHash} and \AMQ interfaces, it is
possible to obtain proofs of certain standard probabilistic properties
or simplifications \emph{for free}.


\begin{figure}[t]
\setlength{\belowcaptionskip}{-10pt}
\centering
\includegraphics[width=0.99\textwidth]{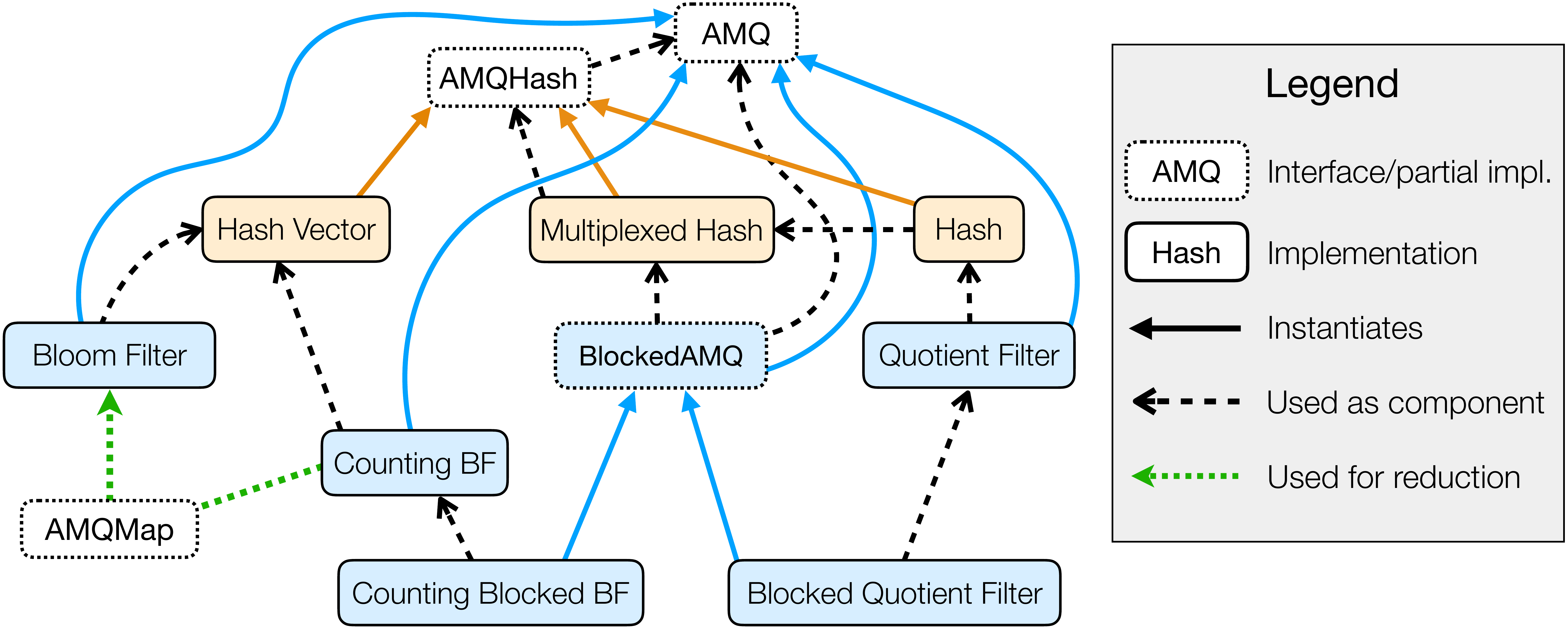}
\caption{Overview of \libname and dependencies the between its
  components.}
\label{fig:overview}
\end{figure}

The diagram in \autoref{fig:overview} provides a high-level overview
of the interfaces of \libname, their specific instances, and
dependencies between them, demonstrating \libname's take on
compositional reasoning and proof reuse.
For instance Bloom filter implementation instantiates the \AMQ
interface implementation and uses, as a component, hash vectors, which
themselves instantiate \lname{AMQHash} used by \AMQ.
Bloom filter itself is also used as a proof reduction target by
Counting Bloom filter.
We will elaborate on this and the other noteworthy dependencies between
interfaces and instances of \libname in the following sections.

\subsection{Counting Bloom Filters through \libname}
\label{sec:count-bloom-filt}

To provide a concrete demonstration of the use of the \AMQ interface,
we now switch over to a new running example---Counting Bloom
filters~\cite{TarkomaRL12}.
A Counting Bloom filter is a variant of the Bloom filter in which
individual bits are replaced with counters, thereby allowing the
removal of elements.
The implementation of the structure closely follows the Bloom filter,
generalising the logic from bits to counters: insertion
increments the counters specified by the hash outputs, while queries
treat counters as set if greater than 0.
In the remainder of this section, we will show how to encode and
verify the Counting Bloom filter for the standard AMQ properties.
We have also proven two novel domain-specific properties of Counting
Bloom filters, which, due to space limits, we outline in
Appendix~\ref{sec:doma-spec-prop}.


First, as the Counting Bloom filter uses the same hashing strategy as
the Bloom filter, the hash interface can be instantiated with the Hash
Vector structure used for the Bloom filter, entirely reusing the
earlier proofs on hash vectors.
Next, in order to instantiate the \AMQ interface, the state parameter
can be defined as a vector of bounded integers, all initially set to
0:
\begin{lstlisting}[mathescape]
Record CF := mkCF { countingbloomfilter_state: m.-tuple $\ordnat{p}$ }.
Definition cf_new : CF := (* a new CF with all counters set to 0 *).
\end{lstlisting}

As mentioned before, the \emph{add} operation increments counters
rather than setting bits, and the \emph{query} operation treats
counters greater than 0 as raised.

\vspace{-10pt}                   %

{\footnotesize{
\begin{align*}
\ccode{cf_add_int} & : \ccode{CF} \rightarrow \ccode{seq} ~  \ordnat{m}  \rightarrow \ccode{CF} \\
\ccode{cf_query_int} & : \ccode{CF} \rightarrow \ccode{seq} ~  \ordnat{m}  \rightarrow \bool
\end{align*}
}}

\vspace{-10pt}

To prevent integer overflows, the counters in the Counting Bloom
filter are bounded to some range $\ordnat{p}$, so the overall data
structure too has a maximum capacity. It would not be possible to
insert any values if doing such would raise any of the counters above
their maximum.
To account for this, the capacity parameter of the AMQ interface is
instantiated with a simple predicate $\ccode{cf_available_capacity}$
that verifies that the structure can support $l$ further inserts by
ensuring that each counter has at least $k * l$ spaces free (where $k$
is the number of hash functions used by the data structure).



The add operation can be shown to be monotone on the value of any
counter when there is sufficient capacity
(Property~\ref{thm:amq-insert-validity}). The remaining properties of
the operations also trivially follow, thereby completing the
instantiation, and allowing the automatic derivation of the
No-False-Negatives result via \autoref{lemma:generalised-enc-fneg}.
%


\subsection{Proofs about False Positive Probabilities by Reduction}
\label{sec:count-bloom-filt}

As the observable behaviour of Counting Bloom filter almost exactly
matches that of the Bloom filter, it seems reasonable that the same
probabilistic bounds should also apply to the data structure.
To facilitate these proof arguments, we provide the \AMQMAP interface
that allows the derivation of probabilistic bounds by reducing one AMQ
data structure to another. 

The \AMQMAP interface is parameterised by two AMQ data structures, AMQ
A and B, using the same hashing operation. It is assumed that
corresponding bounds on False Positive rates have already been proven
for AMQ B, while have not for AMQ A.
The interface first assumes the existence of some mapping from the state
of AMQ A to AMQ B, which satisfies a number of properties:

\begin{lstlisting}
Parameter AMQ_state_map:  A.AMQState -> B.AMQState.
\end{lstlisting}

In the case of our Counting Bloom filter example, this mapping would
convert the Counting Bloom filter state to a bit vector by mapping
each counter to a raised bit if its value is greater than 0.
To provide the of the false positive rate boundary, the \AMQMAP
interface then requires the behaviour of this mapping to satisfy a
number of additional assumptions:

\vspace{3pt}

\begin{prop}[AMQ Mapping Add Commutativity]
  Adding a hash output to the AMQ B obtained by applying the mapping
  to an instance of AMQ A produces the same result as first adding a
  hash output to AMQ A and then applying the mapping to the result.
\end{prop}

\vspace{3pt}

\begin{prop}[AMQ Mapping Query Preservation]
  Applying B's query operation to the result of mapping an instance of
  AMQ A produces the same result as applying A's query operation
  directly.
\end{prop}

\vspace{3pt} 

In the case of reducing Counting Bloom filters (A) to Bloom filters
(B), both results follow from the fact that after incrementing the
some counters, all of them will have values greater than 0 and thus be
mapped to raised bits.

Having instantiated the \AMQMAP interface with the corresponding
function and proofs about it, it is now possible to derive the false
positive rate of Bloom filters for Counting Bloom filters for free
through the following generalised lemma:

\begin{theorem}[AMQ False Positive Reduction]
  For any two AMQs A, B, related by the \emph{\AMQMAP} interface, if
  the false positive rate for B after inserting~$l$ items is given by
  the function $f$ on $l$, then the false positive rate for $A$ is
  also given by $f$ on $l$. That is, in terms of probabilities:
  {\small{
    \begin{gather*}
    \prob{  \cccode{B.AMQ_addm} ~ \xseq ~ (\hshs, \cccode{B.AMQ_new})  \bind  \cccode{B.AMQ_query} ~ y }
    = f (\cccode{length} ~ \xseq) \implies\\
    \prob{  \cccode{A.AMQ_addm} ~ \xseq ~ (\hshs, \cccode{A.AMQ_new})  \bind  \cccode{A.AMQ_query} ~ y }
    = f (\cccode{length} ~ \xseq).
  \end{gather*}
  }}
\end{theorem}


\section{Proof Automation for Probabilistic Sums}
\label{sec:bigsum}

We have, until now, avoided discussing details of how facts about the
probabilistic computations can be composed, and thereby also the
specifics of how our proofs are structured.
As it turns out, most of this process resolves to reasoning about
summations over real values as encoded by \ssr 's \lname{bigop}
library. Our development also relies on the tactic library by
Martin{-}Dorel and Soloviev~\cite{Martin-DorelS16}.

In this section, we outline some of the most essential proof
principles facilitating the proofs-by-rewriting about probabilistic
sums.
While most of the provided rewriting primitives are standalone general
equality facts, some of our proof techniques are better understood as
combining a series of rewritings into a more general rewriting
pattern.
To delineate these two cases, will use the terminology
\textbf{Pattern} to refer to a general pattern our library supports
by means of a dedicated Coq tactic, while \textbf{Lemma} will refer to
standalone proven equalities.

\subsection{The Normal Form for Composed Probabilistic Computations}
\label{sec:relat-prob-comp}

When stating properties on outcomes of a probabilistic computation
(\cf~\autoref{sec:prob-monad}), the computation must first be
recursively evaluated into a distribution, where the intermediate
results are combined using the probabilistic \ccode{bind} operator.
%
%
Therefore, when decomposing a probabilistic property into smaller
subproofs, we must rely on its semantics that is defined for discrete
distributions as follows:
\vspace{-5pt}

{\small{
\[
  \ccode{bind_dist} ~ (P : \dist{A}) ~ (f : A \rightarrow \dist{B}) \eqdef \\
   \sum_{a: ~ A} \sum_{b: ~  B}  P ~ a ~ \times ~ (f ~ a) ~ b
\]
}}

\vspace{-5pt}
\noindent
Expanding this definition, one can represent any statement on the
outcome of a probabilistic computation in a \emph{normal form}
composed of only nested summations over a product of the probabilities
of each intermediate computational step.
This paramount transformation is captured as the following pattern:

\vspace{3pt}

\begin{pattern}[Bind normalisation]
\label{lm:sum-bind}  
{\small{
\begin{gather*}
  \prob{ (c_1 \bind \ldots \bind c_m) = v } = 
  \sum_{v_1} \cdots \sum_{v_{m - 1}} \prob{ c_1 = v_1 } \times \cdots
  \times \prob{ c_m~v_{m-1} = v }
\end{gather*}
}}
\end{pattern}

\noindent
Here, by $c_i~v_{i-1} = v_i$, we denote the event in which the result
of evaluating the command $c_i~v_{i - 1}$ is $v_i$, where $v_{i - 1}$
is the result of evaluating the previous command in the chain. This
transformation then allows us to resolve the proof of a given
probabilistic property into proving simpler statements on its
substeps.
For instance, consider the implementation of Bloom filter's query
operation from Section \ref{sec:prob-monad}.
When proving properties of the result of a particular query (as in
\autoref{lemma:enc-fneg}), we use this rule to decompose the program
into its component parts, namely as being the product of a hash
invocation $\prob{ \ccode{hash_vec_int} ~ x ~ \hshs}$ and the
deterministic query operation $\ccode{bf_query_int}$.
This allows dealing with the hash operation and the deterministic
component \emph{separately} by applying subsequent rewritings to each
factor on the right-hand side of the above equality.



\subsection{Probabilistic Summation Patterns}
\label{sec:comm-prob-proof}

Having resolved a property into our normal form via a tactic
implementing Pattern \ref{lm:sum-bind}, the subsequent reductions rely
on the following patterns and lemmas.

\paragraph{Sequential composition.}
When reasoning about the properties of composite programs, it is
common for some subprogram $e$ to return a probabilistic result that
is then used as the arguments for a probabilistic function $f$.
This composition is encapsulated by the operation $e \bind f$ , as
used by Theorems~\ref{lemma:enc-fneg}, \ref{thm:sing-bit}, and
\ref{lemma:enc-fpos}.
The corresponding programs, once converted to the normal form, are
characterised by having factors within its internal product that
simply evaluate the probability of the final statement
$\ccode{ret} ~ v'$
%
%
to produce a particular value $v_k$:

\vspace{-10pt}

{\small{
\begin{gather*}
  \sum_{v_1} \cdots   \sum_{v_{m - 1}}
\underbrace{
  \prob{ c_1 = v_1 } \times \cdots \prob{ \ccode{ret} ~ v' = v_k }  
  }_{e}
\underbrace{
\cdots \times \prob{ c_m~v_{m - 1} = v }
}_{f}
\end{gather*}
}}

\vspace{-5pt}

\noindent
Since the return operation is defined as a delta distribution with a
peak at the return value $v'$, we can simplify the statement by
removing the summation over $v_k$, and replacing all occurrences of
$v_k$ with $v'$, via the following pattern:
%

%

\vspace{3pt}

\begin{pattern}[Probability of a Sequential Composition]
\label{lm:seqcomp} 
{\small{
\begin{gather*}
  \sum_{v_1} \cdots   \sum_{v_{m - 1}}
   \prob{ \cccode{ret} ~ v' = v_1 }  \cdots \times \prob{ c_m~v_{m-1} = v }] =\\
  \sum_{v_2} \cdots   \sum_{v_{m - 1}}
  \prob{ [v' /v_1 ](c_2~v_1) = v_2 } \times  \cdots \times \prob{ [v'/v_1]c_m~v_{m-1} = v }
\end{gather*}
}}
\end{pattern}
 
\noindent 
Notice that, without loss of generality, Pattern~\ref{lm:seqcomp}
assumes that the $v'$-containing factor is in the head. Our tactic
implicitly rewrites the statement to this form.

\paragraph{Plausible statement sequencing.}
One common issue with the normal form, is that, as each statement is
evaluated over the entirety of its support, some of the dependencies
between statements are obscured. That is, the outputs of one statement
may in fact be constrained to \emph{some subset} of the complete
support.
To recover these dependencies, we provide the following theorem, that
allows reducing computations under the assumption that their inputs
are plausible:

\begin{lemma}[Plausible Sequencing]\label{lm:plausible-sequencing}
  For any computation sequence $c_1 \bind c_2$,
  if it is possible to reduce the computation $c_2 ~ x$ to a simpler
  form $c_3 ~ x$ when $x$ is amongst plausible outcomes of $c_1$,
  (\ie, $\prob{c_1 = x} \neq 0$ holds) then it is possible to rewrite
  $c_2$ to $c_3$ without changing the resulting distribution:
{\small{
\[
  \sum_{x} \sum_{y} \prob{c_1 = x} \times \prob{c_2 ~ x = y} = 
  \sum_{x} \sum_{y} \prob{c_1 = x} \times \prob{c_3 ~ x = y}
\]
}}
\end{lemma}


\paragraph{Plausible outcomes.}
As was demonstrated in the previous paragraph, it is sometimes
possible to gain knowledge that a particular value $v$ is a
plausible outcome for a composite probabilistic computation
$c_1 \bind \dots \bind c_m$:
\begin{gather*}
  \sum_{v_1} \cdots \sum_{v_{m - 1}} \prob{ c_1 = v_1 } \times \cdots \times \prob{ c_m~v_{m-1} = v } \neq 0
\end{gather*}
This fact in itself is not particularly helpful as it does not
immediately provide any usable constraints on the value $v$.
However, we can now turn this inequality into a conjunction of
inequalities for individual probabilities, thus getting more
information about the intermediate steps of the computation:

\begin{pattern}
\label{lm:plausible}
If~
$
  \sum_{v_1} \cdots \sum_{v_{m - 1}} \prob{ c_1 = v_1 } \times \cdots \times \prob{ c_m~v_{m-1} = v } \neq 0,
$
then there exist $v_1, \dots, v_{m-1}$ such that
$
\prob{ c_1 = v_1 } \neq 0 \wedge \cdots  \wedge \prob{ c_m = v } \neq 0.
$
\end{pattern}






\vspace{3pt}

\noindent
This transformation is possible due to the fact that probabilities are
always non-negative, thus if a summation is positive, there must exist
at least one element in the summation that is also positive.

\paragraph{Summary of the development.}
By composing these components together, we obtain a comprehensive
toolbox for effectively reasoning about probabilistic computations.
We find that our summation patterns end up encapsulating most of the
book-keeping associated with our encoding of probabilistic
computations, which, combined with the \AMQ/\lname{AMQHash}
decomposition from \autoref{sec:framework}, allows for a fairly
straightforward approach for verifying properties of AMQs.

\subsection{A Simple Proof of Generalised No False Negatives Theorem}
\label{sec:gen-no-negative}

To showcase the fluid interaction of our proof principles in action,
let us consider the proof of the generalised No-False-Negatives
\autoref{lemma:generalised-enc-fneg}, stating the following:

\vspace{-5pt}

{\small{
\begin{equation}
  \prob{\underbrace{\ccode{AMQ_add} ~ x ~ (\hshs,s)}_{(a), (b)} ~\bind
    ~ \underbrace{\ccode{AMQ_addm} ~ \xseq}_{(c)}~\bind ~
    \underbrace{\ccode{AMQ_query} ~ x}_{(d), (e)} } = 1
\label{eq:nfn}
\end{equation}
}}

\vspace{-5pt}

\noindent
As with most of our probabilistic proofs, we begin by applying
normalisation Pattern~\ref{lm:sum-bind} to reduce the computation into
our normal form:
{\small{
\[
      \sum_{\inds_0, \hshs_0}
      \sum_{s_0}
      \sum_{s_1, \hshs_1}
      \sum_{\inds_2, \hshs_2}
      \left(
\begin{array}{lll}
(a) & \prob{\ccode{AMQHash_hash} ~ x ~ \hshs = (\inds_0, \hshs_0) } & \times \\[2pt]
(b) & \prob{\ccode{ret} ~ (\ccode{AMQ_add_internal}~s~\inds_0) = s_0 }
      & \times \\[2pt]
(c) & \prob{\ccode{AMQ_addm} ~ \xseq ~
           (s_0, \hshs_0) = (s_1,\hshs_1) } & \times \\[2pt]
(d) & \prob{\ccode{AMQHash_hash} ~ x ~ \hshs_1 = (\inds_2, \hshs_2) } &\times \\[2pt]
(e) & \prob{\ccode{ret} ~ (\ccode{AMQ_query_internal} ~ s_1 ~ \inds_2)} 
\end{array}
      \right)
\]
  }}

\noindent
We label the factors to be rewritten as $(a)$--$(e)$ for the
convenience of the presentation, indicating the correspondence to the
components of the statement~\eqref{eq:nfn}.
%
%
From here, as all values are assumed to be unseen, we can use
Property~\ref{thm:amqhash-generalised-uniform} in conjunction with the
sequencing Pattern~\ref{lm:seqcomp} to reduce factors $(a)$ and $(b)$
as follows:
{\small{
    \begin{align*}
      \sum_{\inds_0}
      \sum_{s_1, \hshs_1}
      \sum_{\inds_2, \hshs_2}
      \left(
\begin{array}{lll}
      (a) & \phash & \times \\[2pt]
      (c) & \prob{\ccode{AMQ_addm} ~ \xseq ~ 
           ((s \shortadd \inds_0),
           (\hshs \shorthash (x:\inds_0))) = (s_1,\hshs_1) } & \times \\[2pt]
      (d) & \prob{\ccode{AMQHash_hash} ~ x ~ \hshs_1 = (\inds_2,
            \hshs_2) } & \times \\[2pt]
      (e) & \prob{\ccode{AMQ_query_internal} ~ s_1 ~ \inds_2 }
\end{array}
      \right)
    \end{align*}
  }}
%


\noindent
Here, $\phash$ is the probability from the statement of
Property~\ref{thm:amqhash-generalised-uniform}. We also introduce the
notations $s \shortadd \inds_0$ and $\hshs \shorthash (x:\inds_0)$ to
denote the deterministic  operations $\ccode{AMQ_add_internal}$
and $\ccode{AMQHash_add_internal}$ respectively.
Then, using Pattern~\ref{lm:plausible} for decomposing plausible
outcomes, it is possible to separately show that any plausible
$\hshs_1$ from $\ccode{AMQ_addm}$ must map $x$ to $\inds_0$, as hash
operations preserve mappings.
Combining this fact with Lemma~\ref{lm:plausible-sequencing}
(plausible sequencing) and Hash Consistency
(Property~\ref{thm:general-reducing-hash-vector}), we can derive that
the execution of $\ccode{AMQHash_hash}$ on $x$ in $(d)$ must
return $\inds_0$, simplifying the summation even further:
{\small{
    \begin{align*}
      \sum_{\inds_0}
      \sum_{s_1, \hshs_1}
      \left(
\begin{array}{lll}
      (a) & \phash & \times \\[2pt]
      (c) & \prob{\ccode{AMQ_addm} ~ \xseq ~ 
           ((s \shortadd \inds_0),
           (\hshs \shorthash (x:\inds_0))) = (s_1,\hshs_1) } & \times \\[2pt]
      (e) & \prob{\ccode{AMQ_query_internal} ~ s_1 ~ \inds_0 }
\end{array}
      \right)
    \end{align*}
  }}

Finally, as $s_1$ is a plausible outcome from $\ccode{AMQ_addm}$
called on $s \shortadd \inds_0$, we can then show, using
Property~\ref{thm:amq-query-pres} (query preservation), that querying
for $\inds_0$ on $s_1$ must succeed.
Therefore, the entire summation reduces to the summation of
distributions over their support, which can be trivially shown to be
1.



\section{Overview of the Development and More Case Studies}
\label{sec:casestudies}



\begin{wrapfigure}[11]{r}{0.47\textwidth}
\centering
\vspace{-15pt}
{\scriptsize{
  \begin{tabular}{lcc}
    \toprule 
    Section & \multicolumn{2}{c}{Size (LOC)} \\
            &  Specifications & Proofs  \\
    \midrule
    Bounded containers  & 286 & 1051 \\
    Notation (\S\ref{sec:prob-monad}) &  77 & 0 \\
    Summations (\S\ref{sec:bigsum}) & 742 & 2122 \\
    \midrule
    Hash operations (\S\ref{sec:amqhash-interface}) & 201 & 568 \\
    AMQ framework (\S\ref{sec:amq-interface}) & 594 & 695   \\
    \midrule
    Bloom filter (\S\ref{sec:props})  & 322  & 1088  \\
    Counting BF (\S\ref{sec:count-bloom-filt}, \S\ref{sec:doma-spec-prop}) &  312 & 674 \\
    Quotient filter (\S\ref{sec:quotientfilter}) &  197 & 633 \\
    Blocked AMQ (\S\ref{sec:blockedamq}) &  269 & 522 \\
    \bottomrule    
  \end{tabular}
}}
\end{wrapfigure}
The \libname mechanised framework is implmented as library in Coq
proof assistant~\cite{ceramist}.
It consists of three main sub-parts, each handling a different aspect
of constructing and reasoning about AMQs:
(\emph{i})~a library of \emph{bounded-length data structures},
enhancing \mathcomp's~\cite{Maboubi-Tassi:MathComp} support for
reasoning about finite sequences with varying lengths;
(\emph{ii})~a library of \emph{probabilistic computations}, extending
the \ifth probability theory library~\cite{AffeldtHS14} with
definitions of deeply embedded probabilistic computations and a
collection of tactics and lemmas on summations described in
\autoref{sec:bigsum}; and
(\emph{iii})~the \emph{AMQ interfaces and instances} representing the
core of our framework described in \autoref{sec:framework}.
%
  



Alongside these core components, we also include four specific case
studies to provide concrete examples of how the library can be used for
practical verification.
Our first two case studies are the mechanisation of the Bloom
filter~\cite{Bloom:1970:STH:362686.362692} and the Counting Bloom
filter\cite{TarkomaRL12}, as discussed earlier.
In proving the false-positive rate for Bloom filters, we follow the
proof by Bose \etal\cite{Bose2008Oct}, also providing the first
mechanised proof of the closed expression for Stirling numbers of the
second kind.
Our third case study provides mechanised verification of the quotient
filter\cite{BenderFJKKMMSSZ12}.
Our final case study is a mechanisation of the Blocked AMQ---a family
of 
AMQs with a common aggregation strategy.
We instantiate this abstract structure with each of the prior
AMQs, obtaining, among others, a mechanisation of Blocked Bloom
filters~\cite{PutzeSS09}.
The sizes of each library component, along with the references to the
sections that describe them, are given in the table above.

%




Of particular note, in effect due to the extensive proof reuse
supported by \libname, the proof size for each of our case-studies
\emph{progressively decreases}, with around a 50\% reduction in the
size from our initial proofs of Bloom filters to the final
case-studies of different Blocked AMQs instances.


\subsection{Quotient Filter}
\label{sec:quotientfilter}

A quotient filter~\cite{BenderFJKKMMSSZ12} is a type of AMQ data
structure optimised to be more cache-friendly than other typical AMQs.
In contrast to the relatively simple internal vector-based states of
the Bloom filters, a quotient filter works by internally maintaining a
hash table to track its elements.
%

The internal operations of a quotient filter build upon a fundamental
notion of \emph{quotienting}, whereby a single $p$-bit hash outcome is split
into two by treating the upper $q$-bits (the quotient) and the lower
$r$-bits (the remainder) separately.
Whenever an element is inserted or queried, the item is first hashed
over a single hash function and then the output quotiented.
The operations of the quotient filter then work by using the $q$-bit
quotient to specify a bucket of the hash table, and the $r$-bit
remainder as a proxy for the element, such that a query for an element
will succeed if its remainder can be found in the corresponding
bucket.

A false positive can occur if the outputs of the hash function happen
to exactly collide for two particular values (collisions in just the
quotient or remainder are not sufficient to produce an incorrect
result).
Therefore, it is then possible to reduce the event of a false positive
in a quotient filter to the event that at least one in several draws
from a uniform distribution produces a particular value.
We encode quotient filters by instantiating the \lname{AMQHash}
interface from \autoref{sec:amqhash-interface} with a \emph{single}
hash function, rather than a vector of hash functions, which is used
by the Bloom filter variants (\autoref{sec:overview}).
The size of the output of this hashing operation is defined to be
$2^q * 2^r$, and a corresponding quotienting operation is defined by
taking the quotient and remainder from dividing the hash output
by $2^q$.
With this encoding, we are able to provide a mechanised proof of the
false positive rate for the quotient filter implemented using $p$-bit
hash as being:


\begin{theorem}[Quotient filter False Positive Rate] 
  For a hash-function $\hshs$, after inserting a series of $l$ unseen
  distinct values $\xseq$ into an empty quotient filter
  $\quotientfilter$, for any unseen $y \not\in \xseq$, the probability
  of a query $\bfquery{y}{\quotientfilter}$ for $y$ returning
  \lname{true} is given
  by: 
$\prob{  \cccode{qf_addm} ~ \xseq ~ (\hshs, \cccode{qf_new})  \bind  \cccode{qf_query} ~ y }
    = 1 - \left( 1 - \frac{1}{2^p}\right)^l.$
\label{lemma:qf-enc-fpos}
\end{theorem}
%


%

%
\subsection{Blocked AMQ}
\label{sec:blockedamq}

Blocked Bloom filters\cite{PutzeSS09} are a cache-efficient variant of
Bloom filters where a single instance of the structure is composed of
a vector of $m$ independent Bloom filters, using an additional
``meta''-hash operation to distribute values between the elements.
When querying for a particular element, the meta-hash operation would
first be consulted to select a particular instance to delegate the
query to.

While prior research has only focused on applying this blocking design
to Bloom filters, we found that this strategy is in fact generic over
the choice of AMQ, allowing us to formalise an abstract Blocked AMQ
structure, and later instantiate it for particular choices of
``basic'' AMQs.
As such, this data structure highlights the scalability of \libname
\wrt composition of programs and proofs.

Our encoding of Blocked AMQs within \libname is done via means of two
higher-order modules as in \autoref{fig:overview}: (\emph{i}) a
\emph{multiplexed-hash} component, parameterised over an arbitrary
hashing operation, and (\emph{ii}) a \emph{blocked-state} component,
parameterised over some instantiation of the \AMQ interface.
%
%
%
The multiplexed hash captures the relation between the meta-hash and
the hashing operations of the basic AMQ, randomly multiplexing hashes
to particular hashing operations of the sub-components.
We construct a multiplexed-hash as a composition of the hashing
operation $H$ used by the AMQ in each of the $m$ blocks, and a
meta-hash function to distribute queries between the $m$ blocks.
The state of this structure is defined as pairing of $m$ states of the
hashing operation $H$, one for each of the $m$ blocks of the AMQ, with
the state of the meta-hash function.
As such, hashing a value $v$ with this operation produces a
\emph{pair} of type $(\ordnat{m},\ccode{Value})$, where the first
element is obtained by hashing $v$ over the meta-hash to select a
particular block, and the second element is produced by hashing $v$
again over the hash operation $H$ for this selected block.
With this custom hashing operation, the state component of the Blocked
AMQ is defined as sequence of $m$ states of the AMQ, one for each
block.
The insertion and query operations work on the output of the
multiplexed hash by using the first element to select a particular
element of the sequence, and then use the second element as the value
to be inserted into or queried on this selected state.


Having instantiated the data structure as described above, we proved
the following abstract result about the false positive rate for
blocked AMQs:

\begin{theorem}[Blocked AMQ False Positive Rate]
  For any AMQ $A$ with a false positive rate after inserting $l$
  elements estimated as $f(l)$, for a multiplexed hash-function
  $\hshs$, after having inserted $l$ distinct values $\xseq$, all
  unseen in $\hshs$, into an empty Blocked AMQ filter $\bloomfilter$
  composed of $m$ instances of $A$, for any unseen $y \not\in \xseq$,
  the probability of a subsequent query $\bfquery{y}{\bloomfilter}$
  for $y$ returning \emph{\lname{true}} is given
  by: 
$\prob{  \cccode{BA_addm} ~ \xseq ~ (\hshs, \cccode{BA_new})  \bind  \cccode{BA_query} ~ y }
    = \sum_{i=0}^{l} {l \choose i} (\frac{1}{m})^{i} (1 - \frac{1}{m})^{l - i} f(i).$
\label{lemma:qf-enc-fpos}
\end{theorem}

\noindent
We instantiated this interface with each of the previously defined AMQ
structures, obtaining the Blocked Bloom filters, Counting Blocked
Bloom filters and Blocked Quotient filter along with proofs of similar
properties for them, for free.
%



\newpage

\section{Discussion and Related Work}
\label{sec:related}



\paragraph{Proofs about AMQs.}
\label{sec:analys-amq-struct}

While there has been a wealth of prior research into approximate
membership query structures and their probabilistic bounds, the
prevalence of paper-and-pencil proofs has meant that errors in
analysis have gone unnoticed and propagated throughout the literature.

The most notable example is in Bloom's original
paper\cite{Bloom:1970:STH:362686.362692}, wherein dependencies between
setting bits lead to an incorrect formulation of the bound (equation
(17)), which has since been repeated in several
papers~\cite{Mitzenmacher2002compressed,BroderM03,Dharmapurikar2004Deep,Dharmapurikar2006Longest}
and even textbooks~\cite{mitzenmacher2005}.
While this error was later identified by
Bose~\etal~\cite{Bose2008Oct}, their own analysis was also marred by
an error in their definition of Stirling numbers of the second kind,
resulting in yet another incorrect bound, corrected two years later by
Christensen~\etal~\cite{christensen2010new}, who avoided the error by
eliding Stirling numbers altogether, and deriving the bound directly.
Furthermore, despite these corrections, many subsequent
papers~\cite{PutzeSS09,Jing2009weighted,Li2009memory,debnath2011bloomflash,TarkomaRL12,LimLee2015Complement,QiaoLC11}
still use Bloom's original incorrect bounds.
For example, in Putze~\etal~\cite{PutzeSS09}'s analysis of a Blocked
Bloom filter, they derive an incorrect bound on the false positive
rate by assuming that the false positive of the constituent Bloom
filters are given by Bloom's bound.

\paragraph{Mechanically Verified Probabilistic Algorithms.}
\label{sec:verif-prob-algor}

Past research has also focused on the verification of probabilistic
algorithms, and our work builds on the results and ideas from several
of these developments.



The \alea library also tackles the task of proving properties of
probabilistic algorithms~\cite{audebaud2009proofs}.
In contrast to our choice of a deep embedding for encoding
probabilistic computations, \alea uses a shallow embedding through a
Giry monad~\cite{giry1982categorical}, representing probabilistic
programs as measures over their outcomes.
%
%
As \alea axiomatises a custom type to represent the subset of reals
between 0 and 1 for capturing probabilities, they must independently
prove any properties on reals required for their theorems,
considerably increasing the proof effort.
%

%

%
%

The Foundational Cryptography Framework
(\fcf)~\cite{Petcher-Morissett:POST15} was developed for proving the
security properties of cryptographic programs and provides an encoding
for probabilistic algorithms.
%
Rather than developing specific tooling for solving probabilistic
obligations as we do, their library prioritises a proof strategy of
proving the probabilistic properties of computations by reducing them
to standard ``difficult'' programs with known distributions.
%
%
While this strategy closely follows the typical structure of
cryptographic proofs, their simple encoding increases the complexity
of directly proving probabilistic properties.

Tassarotti~\etal's~\polaris~\cite{tassarotti2019separation} library is
a Coq framework for reasoning about probabilistic concurrent
algorithms.
\polaris uses the same reduction strategy for probabilistic
specifications as the \fcf library, inheriting some of the same issues
with proving standalone bounds.
%
%

H\"{o}lzl considered mechanised verification of probabilistic programs
in Isabelle/ HOL~\cite{Holzl:CPP17}.
While H\"{o}lzl uses a similar composition of probability and
computation monads to encode and evaluate probabilistic programs, his
construction defines the semantics of programs as infinite Markov
chains, represented as a co-inductive stream of probabilistic outputs.
This design makes the encoding unsuitable for capturing terminating
programs, yet it is the only encoding we are aware of that enables
probabilistic proofs about non-terminating programs.

Our previous effort on mechanising the probabilistic properties of
blockchains also considered the encoding of probabilistic computations
in Coq~\cite{Gopinathan-Sergey:CoqPL19}. While that work also relied
on \ifth's probability monad, it primarily considered the
mechanisation of a restricted form of probabilistic properties (those
with complete certainty), and did not deliver reusable tooling for
this task.



While the \libname development is the first, to the best of our
knowledge, that provides a mechanised proof of the probabilistic
properties of Bloom filters, prior research has considered their
deterministic properties.
Blot~\etal~\cite{BlotDL16} provided a mechanised proof of the absence
of false negatives for their implementation of a Bloom filter as part
of their work on a library for using abstract sets to reason about the
bit-manipulations in low-level programs.
%

%


\paragraph{Proofs of differential privacy.}
\label{sec:proofs-diff-priv}

A popular motivation for reasoning about probabilistic computations is
for the purposes of demonstrating differential privacy. 
%

Barthe \etal's \lname{CertiPriv} framework~\cite{BartheKOB12} extends
\alea to support reasoning using a Probabilistic Relational Hoare
logic, and uses this fragment to prove probabilistic non-interference
arguments.
However, \lname{CertiPriv} focuses on proving relational probabilistic
properties of coupled computations rather than explicit numerical
bounds as we do.
More recently, Barthe~\etal~\cite{strub2019relational} have developed
a mechanisation that supports a more general coupling between
distributions.
In the future, we plan to employ \libname for extending the
verification of AMQs to infer the induced probabilistic bounds on
differential privacy guarantees~\cite{ErlingssonPK14}.

%



\section{Conclusion}
\label{sec:conclusion}

The key properties of Approximate Membership Query structures are
inherently probabilistic. Formalisations of those properties are
frequently stated incorrectly, due to the complexity of the underlying
proofs.
We have demonstrated the feasibility of conducting such proofs in a
machine-assisted framework.
The main ingredients of our approach are a principled decomposition of
structure definitions and proof automation for manipulating 
probabilistic sums.
Together, they enable scalable and reusable mechanised proofs about a
wide range of AMQs.

\paragraph{Acknowledgements.~}

We thank Georges Gonthier, Karl Palmskog, George P\^{i}rlea, Prateek
Saxena, and Anton Trunov for their comments on the prelimiary versions
of the paper.
We thank the CPP'20 referees (especially Reviewer D) for pointing out
that the formulation of the closed form for Stirling numbers of the
second kind, which we adopted as an axiom from the work by
Bose~\etal~\cite{Bose2008Oct} who used it in the proof of
\autoref{lemma:enc-fpos}, implied \textsf{False}.
This discovery has forced us to prove the closed form statement in Coq
from the first principles, thus getting rid of the corresponding axiom
and eliminating all potentially erroneous assumptions.
Finally, we are grateful to the CAV'20 reviewers for their feedback.

Ilya Sergey's work has been supported by the grant of Singapore NRF
National Satellite of Excellence in Trustworthy Software Systems
(NSoE-TSS) and by Crystal Centre at NUS School of Computing.










\bibliographystyle{plain}
\bibliography{references}

\appendix
\section{Domain-Specific Properties of Counting Bloom Filters}
\label{sec:doma-spec-prop}

While the No-False-Negatives and false positive rate properties are
practically important aspects of an AMQ, in the case of a Counting
Bloom filter, there are a few other probabilistic behaviours of the
structure that are of importance.
One such property is the ability to remove some elements from a
Counting Bloom filter without affecting queries for other ones, by
decrementing the counters corresponding to the removed element.

To demonstrate the flexibility of our framework, we also provide a
mechanised proof of the validity of this removal operation, which, to
the best of our knowledge, has not been previously formalised:

\begin{theorem}[Counting Bloom filter removal]
  For any Counting Bloom filter $\countingbloomfilter$ with sufficient
  capacity and associated hashes $\hshs$, removing a previously
  inserted value $x'$ will not change the query for any other
  previously inserted value $x$, that is:
$ \prob{
        \cccode{cf_add} ~ x' ~ (\hshs,\countingbloomfilter) \bind
        \cccode{cf_add} ~ x \bind    
        \cccode{cf_remove} ~ x' \bind    
        \cccode{cf_query} ~ x 
      } = 1.$
%
\end{theorem}

The operation $\ccode{cf_remove}$ from the theorem statement deletes a
value from the Counting Bloom filter by decrementing the associated
counters, and is provided as a custom operation externally to the
other \libname components, as removal operations are not a typical
operation in AMQ interfaces.

Our development also provides a proof of another specialised property
of the structure---that inserting any value will increase the total sum
of the counters by a fixed amount.
This property characterises how the modified state of the Counting
Bloom filter allows tracking more detailed information, than just
element membership, in terms of the exact number of insertions.

\begin{theorem}[Certainty of Counter Increments]
\label{thm:counting}
For any counting Bloom filter $\countingbloomfilter$, a value $y$ that
was not previously inserted into $\countingbloomfilter$,
if the sum of the values of all counters $d_i$ in
$\countingbloomfilter$ is $l$, then after inserting $y$, the sum of
the counters will certainly increment by $k$, 
that is:
$\prob{\sum_{d_i \in \countingbloomfilter} d_i = l + k } = 1.$
\end{theorem}


\end{document}
